\documentclass[aps,prb,reprint,superscriptaddress]{revtex4-2}
\usepackage[dvipsnames]{xcolor}
\usepackage{graphicx}
\usepackage{amsmath,amssymb}
\usepackage{braket}
\usepackage{simpler-wick}
\usepackage{makerobust}
\usepackage{hyperref}

\def\equationautorefname~#1\null{%
	Eq.\,(#1)\null
}

\def\figureautorefname~#1\null{%
	Fig.\,#1\null
}

\def\sectionautorefname~#1\null{%
	Sec.\,#1\null
}

\begin{document}

\renewcommand{\vec}[1]{\mathbf{#1}}
\newcommand{\up}{{\uparrow}}
\newcommand{\dw}{{\downarrow}}
\newcommand{\pd}{{\phantom{\dagger}}}
\newcommand{\nn}{\nonumber}
\newcommand{\hc}{\text{H.c.}}
\newcommand{\ie}{\emph{i.e.}}
\newcommand{\eg}{\emph{e.g.}}
\newcommand{\e}[1]{\hat{\vec{e}}_{#1}}

\newcommand{\ketp}[1]{{|#1\rangle_{\rm phys}}}
\newcommand{\ketl}[1]{{|#1\rangle_{\rm log}}}
\newcommand{\braketp}[2]{\langle#1|#2\rangle_{\rm phys}}


\title{Preparation and readout of Majorana qubits\\in magnet-superconductor hybrid systems}

\author{Dan Crawford}
\affiliation{School of Physics, University of Melbourne, Parkville, VIC 3010, Australia}
\author{Roland Wiesendanger}
\affiliation{Department of Physics, University of Hamburg, D-20355 Hamburg, Germany}
\author{Stephan Rachel}
\affiliation{School of Physics, University of Melbourne, Parkville, VIC 3010, Australia}
\noaffiliation

\date{\today}

\begin{abstract}
Initializing the ground state of a quantum bit (qubit) based on Majorana zero modes is one of the most pressing issues for future topological quantum computers. We explore a protocol for initializing such topological qubits based on magnet-superconductor hybrid networks by coupling magnetic chains to a single molecule magnet. The parity of the Majorana state is converted to the presence or absence of a Yu--Shiba--Rusinov state at the molecule. The coupling can be activated by switching the spin state of the molecule, allowing the ground state parity of the chain to be controlled. We demonstrate that initialization with either parity for a Majorana qubit can be achieved. We then introduce the {\it augmented} Majorana qubit, which includes the state of the molecule in the definition of the logical qubit. Using this definition we can initialize a qubit without high-precision timing.
\end{abstract}

\maketitle

{\it Introduction.---} The fragility of quantum states is a significant barrier to realizing large-scale quantum computers. While tremendous effort has resulted in ever increasing coherence times, fault-tolerant architectures may be necessary to overcome noise limitations and to construct large scale quantum computers. By constructing qubits from non-Abelian anyons, inherently fault-tolerant --- or topological --- quantum computing can be achieved\,\cite{nayak_non-abelian_2008}. Unitary operations are implemented by braiding anyons, and measurement by fusion (that is, combining pairs of anyons)\,\cite{nayak_non-abelian_2008,lahtinen_short_2017}. Over the past decade topological superconductivity has emerged as the leading platform for hosting non-Abelian anyons\,\cite{read_paired_2000,kitaev_unpaired_2001}, due to the ability to artificially engineer such systems as nanostructures\,\cite{lutchyn_majorana_2010,oreg_helical_2010,beenakker_search_2020}. Zero-energy subgap modes can be found bound to the ends of one-dimensional topological superconductors; these quasiparticles obey non-Abelian exchange statistics and are called Majorana zero-modes (MZMs)\,\cite{barkeshli_twist_2013}.

While a considerable amount of labor has been invested in semiconductor-superconductor hybrid systems as a route to topological superconductivity\,\cite{lutchyn_majorana_2010,oreg_helical_2010,sau_majorana_2021,microsoft_quantum_inas-hybrid_2023}, magnet-superconductor hybrid (MSH) systems are a compelling alternative\,\cite{nadj-perge_proposal_2013,pientka_topological_2013,nadj-perge_observation_2014,rachel-25pr1}. The state-of-the-art involves assembling chains of magnetic adatoms on the surface of a conventional superconductor by single-atom manipulation using a scanning tunneling microscope tip\,\cite{kim_toward_2018}. In this fashion disorder-free quantum systems can be engineered and analyzed \emph{in situ}\,\cite{schneider-21saeabd7302}. While unambiguous evidence for MZMs has yet to be published, compelling evidence for topological phases in these systems continues to emerge\,\cite{schneider_topological_2021,schneider_precursors_2022,crawford_majorana_2022}.

Upon confirming the presence of MZMs, alongside braiding one of the most pressing issues is qubit initialization. A topological qubit consists of four MZMs (\eg, realized at the ends of two atomic chains), with the logical bit being encoded in the joint parities of \emph{pairs} of MZMs (\ie, the Majorana state). To initialize the qubit from an unknown state thus involves (1) measuring the ground state parity (\ie, the sign of the occupation number of the Majorana state) of each chain, and (2) switching the ground state parity of each chain as required. For semiconductor-superconductor hybrid systems, the premier proposal for state initialization and measurement are dispersive readout protocols\,\cite{smith_dispersive_2020} involving coupled nanowire--quantum dot systems\,\cite{plugge_majorana_2017,karzig_scalable_2017,szechenyi_parity-to-charge_2020}. Such proposals cannot be na\"ively translated to a MSH platform because the MZM wavefunction only penetrates a few angstrom into the substrate\,\cite{crawford_majorana_2022}, so tunneling amplitudes into a nearby quantum dot are strongly suppressed. It may be possible to control the ground state parity of MSH chains by carefully hybridizing the chains to an auxiliary magnetic adatom\,\cite{awoga_controlling_2024}. Here we propose a new protocol, involving a single molecule magnet coupled to a MSH chain. In such a molecule the charge state is intimately related to the magnetic state, so it is relatively straightforward to readout single charges, solving (1). By charging and discharging the molecule it acts as a source/sink of fermions for a MSH qubit, solving (2).

\begin{figure}[t!]
\centering
\includegraphics{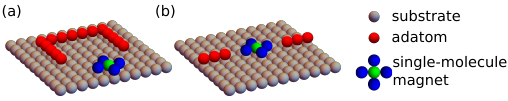}
\caption{\label{fig:setup}
	(a) A Majorana qubit can be initialized by coupling a MSH chain to a single molecule magnet.
	(b) The joint parity of two MSH chains can be read out by coupling both chains to a magnetic molecule.}
\end{figure}

Single molecule magnets (SMMs) have emerged as a powerful tool in spintronics, 2D materials, and other quantum technologies\,\cite{coronado_molecular_2020}. In short, these are coordination complexes involving several ligands bound to a central transition metal. Typically these complexes feature a bistable spin transition, called spin crossover, triggered for example by temperature, light, or pressure\,\cite{gutlich_thermal_1994,real_thermal_2005,molnar_spin_2018}. These complexes can then be exploited as a molecular switch\,\cite{linares_pressure_2012,molnar_spin_2018,kahn_spin-transition_1998}. As with atomic impurities\,\cite{shiba_classical_1968}, a molecular magnet coupled to a superconductor produces Yu-Shiba-Rusinov (YSR) subgap states\,\cite{franke_competition_2011,hatter_magnetic_2015}. Moreover, molecular magnets can also be arranged with single-atom manipulation techniques.

In this Letter we propose a protocol for initializing topological qubits based on MSH networks by coupling magnetic chains to a SMM. The parity of a Majorana state is converted to the presence or absence of a YSR state at the nearby SMM. Coupling between chain and molecule can be activated by switching the spin state of the molecule, allowing the ground state parity of the chain to be controlled. We consider a tight-binding model for this system and find two kinds of Rabi oscillations between chain and molecule, which can be controlled by the Zeeman energy of the molecule, and the molecule-chain and molecule-substrate tunneling amplitudes. We then introduce the augmented Majorana qubit, which includes the state of the molecule in the logical qubit. Using this definition we can initialize a qubit without high-precision timing.

\begin{figure}[t!]
\centering
\includegraphics{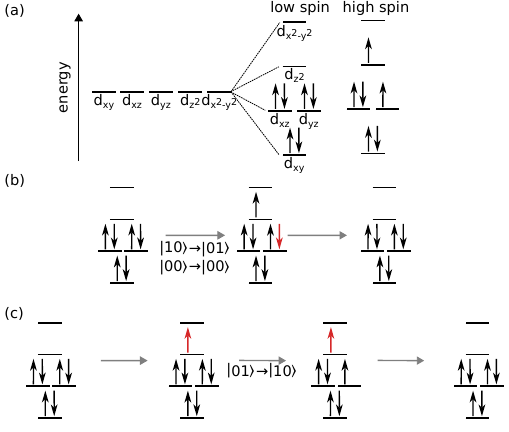}
\caption{\label{fig:protocol}
	(a) Sketch of low spin and high spin states in an iron-based coordination complex. 
	(b) Initialize a chain with unknown parity into the even ground state by (1) switching the magnetic molecule to the high spin state and then (2) if a fermion (red arrow) tunnels from the chain, trap it on the molecule by discharging the molecule. 
	(c) Initialize a chain from the even state into the odd state by (1) exciting the molecule with the STM tip so that (2) a fermion (red arrow) can tunnel from the molecule to the chain and then (3,4) at half a Rabi cycle switching to the low spin state.}
\end{figure}

{\it Setup.---} We consider a chain of magnetic adatoms on a superconducting surface, arranged in a horseshoe such that both ends can be coupled to a magnetic molecule (\autoref{fig:setup}(a)). Note that the ligands in these molecules are not so important, only effecting the potential well the transition metal feels; the low-energy physics is due to the transition metal. We consider magnetic molecules based around $d^6$ transition metals, such as iron(II). In these molecules there are no YSR states in the low spin state while there are up to two YSR states in the high spin state (\autoref{fig:protocol}(a))\,\cite{coronado_molecular_1996:Ch12}. Thus low and high spin states can be distinguished\,\cite{miyamachi_robust_2012,gopakumar_electroninduced_2012} and switched with an STM tip\,\cite{komeda_observation_2011,miyamachi_robust_2012,gopakumar_electroninduced_2012}. The origin of these states can be understood from crystal field theory\,\cite{ballhausen_introduction_1962,figgis_ligand_2000} as either classical or quantum spins\,\cite{von_oppen_yu-shiba-rusinov_2021}. We assume that there is an orbital (\eg, $d_{yz}$) close to the Fermi energy so it may couple to the Majorana state on the chain --- this orbital must be unpaired in the high spin state. We assume that there is no tunneling into other unoccupied orbitals, which may be true if the energy barrier is large enough and the thermal energy of the system low enough\,\cite{mugarza_electronic_2012, cantalupo_highspin_2012}. For simplicity here we describe the molecule as being charge neutral, but strictly speaking this is not necessary.

Consider a molecule coupled to a chain in the even parity ground state. Because the transition metal in the molecule is positively charged it is a Lewis acid, \emph{i.e.}, it is an electron accepter (this remains true for a SMM coupled to a substrate). Hence, regardless of the spin state, tunneling from the molecule to the chain is suppressed. Thus by cycling the molecule from low spin, to high spin, and then back to low spin the ground state parity of the chain does not change. Simultaneously, the number of YSR states on the molecule will cycle from zero, to two, and then back to zero.

However, when the chain is in the odd parity ground state, and the molecule is in the high spin state, a fermion can tunnel from the chain into the molecule. If this happens the SMM becomes charged and transitions to an excited state\,\cite{meded_electrical_2011,komeda_observation_2011,bartolome_molecular_2014:Ch9}. The transition metal then becomes a weaker Lewis acid --- the more positively charged an atom is, the stronger a Lewis acid it is. Thus the fermion, initially on the chain, can tunnel between the chain and molecule. Because this is essentially a two-level problem we expect that there will be Rabi oscillations between the chain and molecule. These Rabi oscillations could be observed by high-resolution STM measurements\,\cite{liang_ultrafast_2023} --- essentially by observing a time-dependent YSR peak in the spectra --- or via electron spin resonance\,\cite{willke_coherent_2021,sellies_single-molecule_2023}. By discharging the molecule at half a Rabi cycle it could be possible to ``trap" the oscillating fermion on the molecule. Thus when the chain is in an unknown ground state we can initialize it into the even parity state with the following protocol: (1) Initialize the SMM into the low spin state. (2) Cycle the SMM into the high spin state. (3a) Check if the SMM is charged (\ie, the number of YSR states); if it is, discharge the SMM at half a Rabi cycle, which returns the SMM to the low spin state. (3b) If the SMM is not charged, cycle the SMM back to the low spin state. This protocol is sketched in \autoref{fig:protocol}(b).

Suppose we have successfully initialized the chain in the even parity ground state. We can then switch the ground state parity of the chain by introducing an additional electron into the system. This can be achieved by charging the molecule with the STM tip. The STM tip induces a local gate
to the molecule\,\cite{pham_selective_2019}, such that an electron on the STM tip can tunnel into one of the higher energy orbitals\,\cite{krull_site-_2013,steurer_manipulation_2015}. This changes the (Lewis) acidity of the molecule, which could allow for a fermion to tunnel from the molecule into the chain. We then expect Rabi oscillations between the chain and molecule. Thus when the chain is in the even parity ground state we can initialize into the odd parity state with the following protocol: (1) Initialize the SMM into the low spin state. (2) Charge the SMM with an electron. (3) The SMM will Rabi oscillate between excited low spin\,\cite{meded_electrical_2011,bartolome_molecular_2014:Ch9} and high spin states. (4) Switch the SMM at half a Rabi cycle from excited low spin to low spin. This protocol is sketch in \autoref{fig:protocol}(c).

{\it Model.---} We construct a tight-binding model on a two-dimensional square lattice spanned by $\e{1}$ and $\e{2}$. We model the superconducting substrate as well as the magnetic adatoms and the molecule as separate sites hosting spinful electrons.
Superconductivity is not directly applied to the magnetic adatoms and molecule, but only through the coupling to the superconducting substrate effectively induced.
The adatoms are restricted to sites $\Lambda$ and the magnetic molecule to site $\vec{R}$. The ends of the chain of adatom are sites $\Lambda'$. The Hamiltonian is $H = H_{\rm sub} + H_{\rm adatom} + H_{\rm mol} + H_{\rm \Gamma}$ with the substrate contribution,
\begin{align}
\nn
&H_{\rm sub} = \sum_\vec{r} \bigg[ t \left( c_\vec{r}^\dagger c_{\vec{r} + \e{1}}^\pd + c_\vec{r}^\dagger c_{\vec{r} + \e{2}}^\pd \right) - \frac{\mu_{\rm sub}}{2} c_\vec{r}^\dagger c_\vec{r}^\pd + \Delta c_{\vec{r},\up}^\dagger c_{\vec{r},\dw}^\dagger
\\
&\qquad + i\alpha \left( c_\vec{r}^\dagger \sigma_2 c_{\vec{r}+\e{1}}^\pd - c_\vec{r}^\dagger \sigma_1 c_{\vec{r}+\e{2}}^\pd \right) + \hc \bigg],
\end{align}
the terms describing the adatom chain,
\begin{align}
\nn
&H_{\rm adatom} = \sum_{\vec{r} \in \Lambda} \bigg[ t \left( a_\vec{r}^\dagger a_{\vec{r} + \e{1}}^\pd + a_\vec{r}^\dagger a_{\vec{r} + \e{2}}^\pd \right) - \mu_{\rm adatom} a_\vec{r}^\dagger a_\vec{r}^\pd\\
&\qquad + J a_\vec{r}^\dagger \sigma_3 a_\vec{r}^\pd + \hc \bigg],
\end{align}
the SMM contribution,
\begin{align}
&H_{\rm mol} = -\mu_{\rm mol} b_\vec{R}^\dagger b_\vec{R}^\pd + J_{\rm mol} b_\vec{R}^\dagger \sigma_3 b_\vec{R}^\pd,
\end{align}
and finally the coupling between SMM/adatom and substrate,
\begin{align}
\nn
&H_{\rm \Gamma} = \sum_{\vec{r} \in \Lambda} t \left( a_\vec{r}^\dagger c_\vec{r}^\pd + a_\vec{r}^\dagger c_{\vec{r}+\e{1}} + a_\vec{r}^\dagger c_{\vec{r}+\e{2}} + a_\vec{r}^\dagger c_{\vec{r}+\e{1}+\e{2}} + \hc \right) \\
\nn
&~ + \Gamma (b_\vec{R}^\dagger c_\vec{R}^\pd + b_\vec{R}^\dagger c_{\vec{R}+\e{1}} + b_\vec{R}^\dagger c_{\vec{R}+\e{2}} + b_\vec{R}^\dagger c_{\vec{R}+\e{1}+\e{2}} + \hc )\\
&~ + V\sum_{\vec{r} \in \Lambda'} \left( b_\vec{R}^\dagger a_{\vec{r}} + \hc \right) .
\end{align}
Here $c_\vec{r}^\dagger = (c_{\vec{r},\up}^\dagger, c_{\vec{r},\dw}^\dag)$ is a spinor of the creation operators for electrons on the substrate at site $\vec{r}$ with spin $\up,\dw$. $a^\dagger$ and $b^\dagger$ are spinors for electrons on the adatoms and the magnetic molecule respectively. $\sigma_{1,2,3}$ are the three Pauli matrices. $t$ is the adatom-adatom, superconductor-superconductor, and adatom-superconductor hopping parameter. While in general these hoppings should have different values, we choose a common energy to reduce the number of free parameters. We find qualitatively the same results by setting the adatom hoppings to be smaller than the superconductor hoppings. $\mu_{\rm sub}$ is the chemical potential on the substrate; $\mu_{\rm adatom}$ the chemical potential on the adatoms; $\alpha$ the Rashba spin-orbit coupling strength; $\Delta$ the $s$-wave superconducting amplitude; $J$ the Zeeman strength resulting from the magnetic moments of the adatoms; $\mu_{\rm mol}$ the chemical potential of the molecule; $J_{\rm mol}$ the Zeeman strength of the molecule; $\Gamma$ the coupling between the molecule and the substrate; and $V$ the coupling between the molecule and the adatoms. $\Gamma$ and $V$ could be tuned by choosing the appropriate molecule and molecule-chain separation.

\begin{figure}[t!]
\centering
\includegraphics{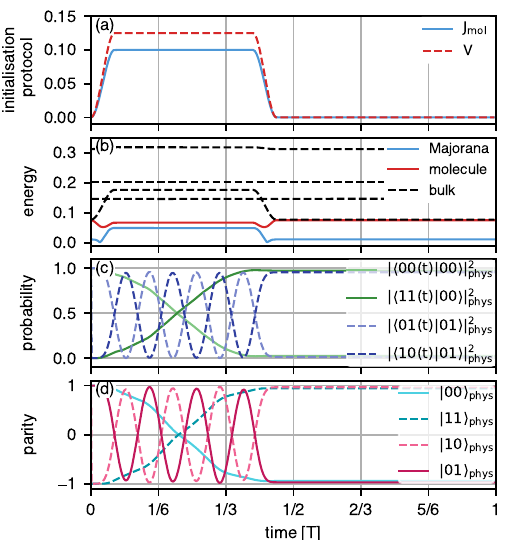}
\caption{\label{fig:results}
	(a) Initialization protocol. $J_{\rm mol}$ and $V$ are switched on rapidly.
	(b) Instantaneous energies under the initialization protocol. Energies of the magnetic molecule and MZM are highlighted. Negative energies not shown for conciseness.
	(c) Same-state and non-vanishing different-state transition probabilities for each of the four relevant states.
	(d) Parity of the chain $\langle \mathcal{P} \rangle$ for each of the four relevant states.
	Parameters: $(\mu_{\rm sub}, \mu_{\rm adatom}, \mu_{\rm mol}, J, \alpha, \Delta) = (2, -1.666, 0.075, -1, 0.25, 0.5)t$, $(N_x, N_y, L_1, L_2) = (13, 10, 10, 6)$, $(T, \delta t) = (470 \hbar/t, 0.5)$
}
\end{figure}

We leave detailed modeling of spin crossover to future work and instead approximate the protocol described above by sweeping $V,J_{\rm mol}$ from zero (low spin) to some maximum $V^{\rm max}, J_{\rm mol}^{\rm max}$ (high spin), and then back to zero as shown in \autoref{fig:results}(a).  We label our physical Fock states $\ket{\vec{n}_d(t)} = \ketp{n_1n_2\dots}$ with $n_x \in [0, 1]$ such that $n_1$ is the occupancy of the ground state of the chain-molecule-superconductor hybrid system, $n_2$ the occupancy of the first excited state, and so on. As discussed above we model the molecule as a single orbital, and choose parameters so that the state associated with the molecule is a subgap state. Thus $n_1$ is the occupancy of the Majorana state on the chain, and $n_2$ is the occupancy of the molecule.
For simplicity we consider only the states $\{\ketp{00}, \ketp{10}, \ketp{01}, \ketp{11}\}$. Using the time-dependent Pfaffian method\,\cite{mascot_many-body_2023} we compute the fidelities $|\braketp{n}{n'}|^2$ of these four states, as well as the ground state parity of the chain $\langle \mathcal{P} \rangle = 1 - 2d_1^\dagger d_1^\pd$. $V^{\rm max}, J_{\rm mol}^{\rm max}$ are chosen so that when the chain and molecule are coupled, the Majorana and molecule states are close in energy (\autoref{fig:results}(b)). Due to the finite size of the system the two MZMs hybridize slightly, resulting in a small but finite energy for the Majorana state.

Consider the states $\{\ketp{01}, \ketp{10} \}$.
When the coupling is switched on a fermion can tunnel between the chain and molecule. As expected, this leads to Rabi oscillations between the states $\ketp{01}$ and $\ketp{10}$. At half a Rabi cycle the chain and molecule are decoupled, and the system which started in $\ketp{01}$ transitions to $\ketp{10}$ (\autoref{fig:results}(c)). Or in other words, the ground state parity of the chain flips from even to odd (\autoref{fig:results}(d)).

Now consider the states $\{\ketp{00}, \ketp{11} \}$.
Surprisingly, we see when the coupling is switched on there are transitions between $\ketp{00}$ and $\ketp{11}$ (\autoref{fig:results}(c)). These are complemented by oscillations in the parity of the chain (\autoref{fig:results}(d)). By performing some numerical experiments\,\cite{supp} we find that the Rabi frequency is controlled by the gap size and the energy of the MZMs. We complement these results by analyzing a minimal model --- a two site Kitaev chain coupled to an impurity --- in which we find two characteristic Rabi oscillations, exactly corresponding to those shown in \autoref{fig:results}. The transition probabilities are $P_{1, 2} \approx A_{1, 2} \sin^2 \Omega_{1, 2}$. In general, $\Omega_1 \neq \Omega_2$ and $A_1 \neq A_2$ \cite{supp}. Assuming a hopping strength of 100 meV, these Rabi cycles are of the order $0.1 - 10$ ns. Time-resolutions of about 1 ns is routinely achievable by STM via pump-probe techniques \cite{friedlein_radio-frequency_2019}.



\begin{figure}[t!]
\centering
\includegraphics{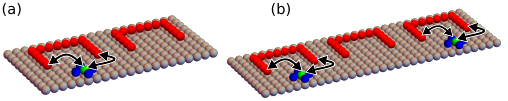}
\caption{\label{fig:dressed-majorana-qubit}
	(a) Augmented Majorana qubit comprised of two chains. One chain is coupled to a magnetic molecule.
	(b) Two augmented Majorana qubits.
}
\end{figure}

{\it Augmented Majorana qubit.---} If it were possible to switch the spin state of a SMM at precisely half a Rabi cycle, one could use the above Rabi oscillations to control the ground state parity of a MSH chain.
Recent advances --- such as improved spectral resolution\,\cite{van_weerdenburg_scanning_2021} and the development of spin-polarised techniques\,\cite{wiesendanger_spin_2009,willke_probing_2018,baumann_electron_2015,bae_enhanced_2018} --- enable the study of the spin dynamics of single atoms\,\cite{yang_coherent_2019}, few-atom quantum magnets\,\cite{khajetoorians-13s55} and single molecules\,\cite{willke_coherent_2021,sellies_single-molecule_2023}.
This opens up an intriguing possibility: to include the occupation of a magnetic molecule in the definition of a Majorana qubit, which then allows one to distinguish logical states by their characteristic Rabi frequencies. We propose a simple change to the conventional Majorana qubit: add a SMM, coupled to one of the two $p$-wave chains (\autoref{fig:dressed-majorana-qubit}(a)), and consider the occupation of the SMM as well as the two chains. The qubit is then defined by \emph{sets} of physical states, \eg,
\begin{align}
&\ketl{0} = \{ \ketp{n_100}, \ketp{n_111} \}, \\
&\ketl{1} = \{\ketp{n_110}, \ketp{n_101}\}.
\end{align}
Here for $\ketp{n_1n_2n_3}$, there is a degenerate ground state due to Majorana states on both chains, and a subgap state due to the molecule. $n_1$ is the ground state occupancy of one of the chains, $n_2$ the ground state occupancy of the chain which directly couples to the molecule, and $n_3$ the occupancy of the molecule. As above, we model the molecule as a single orbital.
Similar concepts were introduced in Refs.\,\cite{akhmerov_topological_2010} and \cite{cheng_nonadiabatic_2011}. These works consider qubits consisting of vortices binding Majorana bound states as well as additional excitations. Both the Majorana bound state and additional excitations are involved in braiding, because all states are localized to the same vortex. Ref.\,\cite{smith_dispersive_2020} also discusses a Majorana qubit initialisation technique via a quantum dot. This similarly results in a dressed qubit where the state of the quantum dot is involved in initialisation and braiding. In contrast, here the SMM state augments the Majorana qubit and is only involved in initialisation and readout, not braiding. We thus call this an \emph{augmented} Majorana qubit.

Such a qubit can be initialised without measuring the occupation of the chains. From the results above we have that oscillations between $\ketp{n_1 1 0}$ and $\ketp{n_1 0 1}$ have frequency $\Omega_1$, while oscillations between $\ketp{n_1 0 0}$ and $\ketp{n_1 1 1}$ have frequency $\Omega_1$. That is, we can distinguish the ground states $\{\ketp{n_100},\ketp{n_111}\}$ from $\{\ketp{n_110},\ketp{n_101}\}$ by their characteristic Rabi frequencies. This means we can define $\ketl{0}$ as \emph{either} of the physical states $\{\ketp{n_100},\ketp{n_111}\}$, and $\ketl{1}$ as either of the physical states $\{\ketp{n_100},\ketp{n_111}\}$.

By combining the protocols shown in \autoref{fig:protocol}(b,c) we can initialize the qubit: (1) Start with the molecule in the low spin state. (2) Switch chain-molecule couplings on by switching to the high spin state. Measure a frequency $\Omega_1$ as $\ketp{n_1n_2n_3}$ oscillates into $\ketp{n_1n_2'n_3'}$. (3) Switch chain-molecule couplings off. Charge the molecule. (4) Switch chain-molecule couplings on. Measure a frequency $\Omega_2$ as $\ketp{n_1n_2n_3}$ oscillates into $\ketp{n_1n_2''n_3''}$. (5) Switch couplings off, returning the molecule to the low spin state. We then define the logical qubit $\ketl{0}$ as either of the physical states associated with $\Omega_1$, and $\ketl{1}$ as either of the states associated with $\Omega_2$. Note that at no point in this protocol do we explicitly measure the ground state parity of either chain, nor does the SMM need to be switched with high-precision timing. It is straightforward to apply this protocol to multiple qubits (\autoref{fig:dressed-majorana-qubit}(b)).

By explicit computation\,\cite{supp} we find that that these qubit definitions preserve the exact same Clifford gates as used for a conventional Majorana qubit. Thus all Clifford gates can be realized by the usual braids\,\cite{lahtinen_short_2017}.


{\it Outlook.---} This protocol has the following requirements for the molecular magnet (1) a low/high-spin electronic structure as sketched in \autoref{fig:protocol}; (2) a YSR state close to the Fermi energy and in the topological gap, which may be less than $200 \mu$eV \cite{schneider_topological_2021}; and (3) this YSR state must be distributed and oriented in a way such that it can couple to a MSH chain. We suggest that the best candidate molecules are the metal phthalocyanines and metal porphyrins. It has been established that these molecules can be deposited on superconductors\,\cite{franke_competition_2011,hatter_magnetic_2015,kezilebieke_coupled_2018,kezilebieke_observation_2019,rubio-verdu_coupled_2021,li_on-surface_2022, xia_spin-orbital_2022,shahed_observation_2022}, and manipulated with an STM tip\,\cite{franke_competition_2011,xia_spin-orbital_2022} to create dimers\,\cite{kezilebieke_coupled_2018} and clusters\,\cite{amokrane_role_2017,li_on-surface_2022}. YSR states close to the Fermi energy\,\cite{hatter_magnetic_2015,kezilebieke_coupled_2018,kezilebieke_observation_2019,rubio-verdu_coupled_2021,li_on-surface_2022,xia_spin-orbital_2022,shahed_observation_2022} are observed in these hybrid superconductor--molecular magnet systems, and in many cases there is significant intermolecular coupling\,\cite{mugarza_electronic_2012,amokrane_role_2017,kezilebieke_coupled_2018,shahed_observation_2022}.
An STM or force microscope tip can be used to both switch a molecule\,\cite{komeda_observation_2011,miyamachi_robust_2012,gopakumar_electroninduced_2012,weiser-13njp013011} and to control the charge state\,\cite{martinez-blanco_gating_2015,pham_selective_2019,steurer-15nc8353}. This protocol has no specific temperature requirements; already MSH experiments operate at millikelvin temperatures \cite{schneider_topological_2021,schneider-21saeabd7302,schneider_precursors_2022}, minimizing thermal broadening. However, even at low temperatures quasiparticles can be thermally excited. In general, these quasiparticles can decohere Majorana qubits (this is known as quasiparticle poisoning). For simplicity, the protocols introduced here explicitly assume that there is no quasiparticle poisoning. Additional measures are required to either suppress quasiparticle poisoning, or to correct for quasiparticle errors via standard quantum error correction.

It should be clear that the initialisation protocols discussed here are quantum demolitional. This is because we explicitly rely on controlled quasiparticle transport into the MZM states. As a consequence, our approach cannot be used for measurement-only techniques\,\cite{bonderson_measurement-only_2008}. We also point out that MZMs might appear in chains of magnetic molecules, as long as there is sufficient hybridization between molecules. One could then tune parts of the chain from the topological to the trivial phase by switching off magnetism molecule-by-molecule. This could be an alternative route to braiding.


\begin{acknowledgments}
The authors acknowledge discussions with Eric Mascot.
S.\ R.\ acknowledges support from the  Australian Research Council through Grants No.\ DP200101118 and DP240100168.
R.\ W.\ acknowledges funding of the
European Union via the ERC Advanced Grant (BE) 786020
and by the Cluster of Excellence “Advanced Imaging of Mat-
ter” of the Deutsche Forschungsgemeinschaft (DFG) (DE),
EXC 2056, 390715994.
This research was supported by The University of Melbourne’s Research Computing Services and the Petascale Campus Initiative.
 This research was undertaken with the use of the National Computational Infrastructure (NCI Australia). NCI Australia is enabled by the National Collaborative Research Infrastructure Strategy (NCRIS). 
\end{acknowledgments}

\bibliography{parity-to-magnetism}

\end{document}